\documentclass[twocolumn,preprintnumbers,aps,pra,superscriptaddress,amsmath,amssymb]{revtex4-1}

\usepackage{graphicx}
\usepackage{dcolumn}
\usepackage{bm}
\usepackage{bbm}
\usepackage{hyperref}
\usepackage{color}
\usepackage{multirow}
\usepackage{physics}

\begin{document}

\title{Proximity-induced artefacts in magnetic imaging with nitrogen-vacancy ensembles in diamond}

\author{J.-P. Tetienne} 
\email{jtetienne@unimelb.edu.au}
\affiliation{School of Physics, The University of Melbourne, VIC 3010, Australia}	
	
\author{D. A. Broadway}
\affiliation{School of Physics, The University of Melbourne, VIC 3010, Australia}	
\affiliation{Centre for Quantum Computation and Communication Technology, School of Physics, The University of Melbourne, VIC 3010, Australia}

\author{S. E. Lillie}
\affiliation{School of Physics, The University of Melbourne, VIC 3010, Australia}	
\affiliation{Centre for Quantum Computation and Communication Technology, School of Physics, The University of Melbourne, VIC 3010, Australia}

\author{N. Dontschuk}
\affiliation{Centre for Quantum Computation and Communication Technology, School of Physics, The University of Melbourne, VIC 3010, Australia}

\author{T. Teraji}
\affiliation{National Institute for Materials Science, Tsukuba, Ibaraki 305-0044, Japan}

\author{L. T. Hall}
\affiliation{School of Physics, The University of Melbourne, VIC 3010, Australia}

\author{A. Stacey}
\affiliation{Centre for Quantum Computation and Communication Technology, School of Physics, The University of Melbourne, VIC 3010, Australia}

\author{D. A. Simpson}
\affiliation{School of Physics, The University of Melbourne, VIC 3010, Australia}

\author{L. C. L. Hollenberg}
\affiliation{School of Physics, The University of Melbourne, VIC 3010, Australia}
\affiliation{Centre for Quantum Computation and Communication Technology, School of Physics, The University of Melbourne, VIC 3010, Australia}

\date{\today}
	
\begin{abstract}
	
Magnetic imaging with ensembles of nitrogen-vacancy (NV) centres in diamond is a recently developed technique that allows for quantitative vector field mapping. Here we uncover a source of artefacts in the measured magnetic field in situations where the magnetic sample is placed in close proximity (a few tens of nm) to the NV sensing layer. Using magnetic nanoparticles as a test sample, we find that the measured field deviates significantly from the calculated field, in shape, amplitude and even in sign. By modelling the full measurement process, we show that these discrepancies are caused by the limited measurement range of NV sensors combined with the finite spatial resolution of the optical readout. We numerically investigate the role of the stand-off distance to identify an artefact-free regime, and discuss an application to ultrathin materials. This work provides a guide to predict and mitigate proximity-induced artefacts that can arise in NV-based wide-field magnetic imaging, and also demonstrates that the sensitivity of these artefacts to the sample can make them a useful tool for magnetic characterisation.

\end{abstract}

\maketitle

\section{Introduction}

The nitrogen-vacancy (NV) centre in diamond is a point defect that can be used as an atomic-sized sensor by exploiting the properties of its quantum spin \cite{Doherty2013,Rondin2014,Schirhagl2014}. Among its attractive features is the variety of physical quantities it can measure (magnetic field \cite{Taylor2008,Degen2008}, electric field \cite{Dolde2011}, temperature \cite{Kucsko2013} etc.) as well as the different modes of operation avaiable (DC \cite{Balasubramanian2008} or AC \cite{Maze2008} field sensing, noise sensing \cite{Cole2009}). In this work, we focus primirily on the NV centre operated as a DC magnetometer, which relies on measuring Zeeman shifts of the spin sublevels via optically detected magnetic resonance (ODMR) \cite{Rondin2014}. Using a single NV centre, magnetic sensitivities under 1 $\mu$T/Hz$^{1/2}$ have been demonstrated, with a probe volume of about (1 nm)$^3$ given by the size of the defect \cite{Balasubramanian2009,Dreau2011}. To form an image of the magnetic field produced by a sample, a commonly employed approach is to scan a single NV centre above the magnetic sample \cite{Balasubramanian2008,Maletinsky2011,Rondin2012,Tetienne2014,Pelliccione2016}. The spatial resolution is then limited by the NV-sample distance and can be as low as 10 nm \cite{Thiel2016}. However, this approach is inherently slow and technically challenging. Another approach involves creating a quasi-two-dimensional ensemble of NV centres near the diamond surface, placing the magnetic sample directly on the diamond, and performing wide-field ODMR spectroscopy of the NV layer using a camera \cite{Steinert2010,Pham2011,Chipaux2015,Simpson2016}. This method provides faster image acquisition, and enables full vector magnetic field mapping over relatively large fields of view (typically $100\times100~\mu$m$^2$), with a spatial resolution ultimately limited by the diffraction of light ($\approx350$~nm). In the past few years, this approach has been applied to a remarkably diverse range of topics, from imaging magnetism in biological and geological samples \cite{LeSage2013,Fu2014,Glenn2015}, to mapping electrical currents in graphene \cite{Tetienne2017}. In this work, we uncover a previously unrecognised and potentially significant source of artefacts in the measured magnetic field maps, which can occur when a ferromagnetic sample is placed in close proximity ($\lesssim200$~nm) to the NV layer. We investigate this effect via a combination of experiments, using magnetic nanoparticles deposited on the diamond, and numerical simulations. Ways to mitigate these artefacts, or on the contrary to use them as a resource, are discussed. 

\section{Experiment}

\begin{figure*}[t!]
	\begin{center}
		\includegraphics[width=0.99\textwidth]{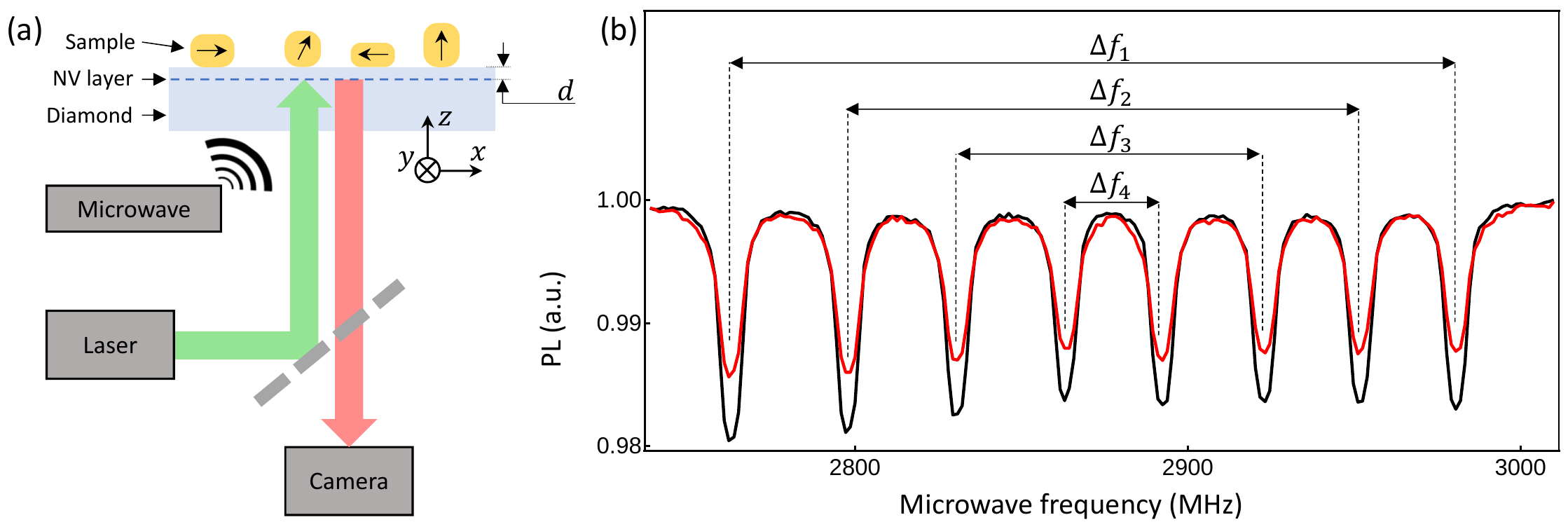}
		\caption{(a) Schematic of the experimental setup. The sensing platform consists of a diamond substrate hosting a layer of near-surface nitrogen-vacancy (NV) centres. The magnetic samples to be imaged (here magnetic nanoparticles shown as yellow objects; the arrows depict their magnetisation) are placed directly on the diamond surface. The NV photoluminescence (PL) under green laser and microwave excitations is imaged on a camera to form the magnetic field image of the sample via optically detected magnetic resonance (ODMR). (b) Example ODMR spectra obtained from a single imaging pixel $500\times500$~nm$^2$ in size. Measurement of the four Zeeman splittings $\Delta f_i$ ($i=1...4$) allows reconstruction of the local vector magnetic field. The two spectra correspond to two different locations on the sample, namely near (red) and laterally far from (black) a magnetic particle (see details in text and Fig. \ref{Fig2}c).}
		\label{Fig1}
	\end{center}
\end{figure*} 

The imaging setup considered in this work is depicted in Fig. \ref{Fig1}a. The sensing element is a diamond substrate with a near-surface layer of NV centres created at a distance $d$ below the top surface, which hosts the magnetic sample to be imaged. The red photoluminescence (PL) emitted by the NV centres under illumination by a green laser is imaged on a camera, and modulated with a microwave source to obtain an ODMR spectrum at each imaging pixel. Fig. \ref{Fig1}b shows typical ODMR spectra acquired under conditions optimised for fast vector field mapping, where a small bias magnetic field ${\bf B}_0$ (here of amplitude $B_0=4.4$~mT) is aligned such that the projections onto the four different NV orientations (corresponding to the four [111] diamond crystal axes) are non-zero and unequal, resulting in eight separated resonance lines \cite{Steinert2010,Chipaux2015,Glenn2017}. Due to the strong crystal field, the Zeeman splitting $\Delta f_i$ of each pair of lines ($i=1$ to 4, as defined in Fig. \ref{Fig1}b) is proportional, to a good approximation, to the magnetic field projection along the corresponding NV symmetry axis \cite{Rondin2014}. Therefore, it is possible to deduce the vector components of the local magnetic field in the lab frame (depicted in Fig. \ref{Fig1}a, $z$ being normal to the diamond surface), knowing the relative orientation of the diamond crystal \cite{Steinert2010,Maertz2010,Chipaux2015,Tetienne2017}. 

As a test sample, we used FeNiCr nanoparticles deposited directly on the diamond surface. The particles feature a range of sizes from 10 to several 100s of nanometers (as measured by atomic force microscopy), and are expected to form a single ferromagnetic domain. The diamond used for this work is a 30-$\mu$m-thick slab overgrown with 2 $\mu$m of $^{12}$C-enriched diamond via chemical vapour deposition \cite{Teraji2015}. It was implanted with nitrogen ($^{14}$N$^+$) ions at a dose of $10^{13}$ ions/cm$^2$ and energy 4 keV, and annealed at 1200$^\circ$C to form NV centres \cite{Tetienne2018}. According to previous simulations and measurements \cite{FavarodeOliveira2015,Lehtinen2016,DeOliveira2017,Tetienne2018}, the NV centres are expected to be distributed up to 20 nm from the surface. We performed magnetic imaging of the magnetic nanoparticles using ODMR measurements in the same conditions as in Fig. \ref{Fig1}b, i.e. with a bias magnetic field ${\bf B}_0=(0.68,-2.08,3.80)$~mT. For each pixel of the camera, the ODMR spectrum is fitted with a sum of eight Lorentzian lines, from which we extract the four splittings $\Delta f_i$. Knowing the direction of ${\bf B}_0$ relative to each NV axis, the magnetic field components are then simply given by $B_x=(\Delta f_2-\Delta f_3)/\gamma_e\sqrt{32/3}$, $B_y=(\Delta f_4-\Delta f_1)/\gamma_e\sqrt{32/3}$ and $B_z=(\Delta f_2+\Delta f_3)/\gamma_e\sqrt{16/3}$, where $\gamma_e=28$ GHz/T is the electron gyromagnetic ratio. 

\begin{figure*}[t!]
	\begin{center}
		\includegraphics[width=0.8\textwidth]{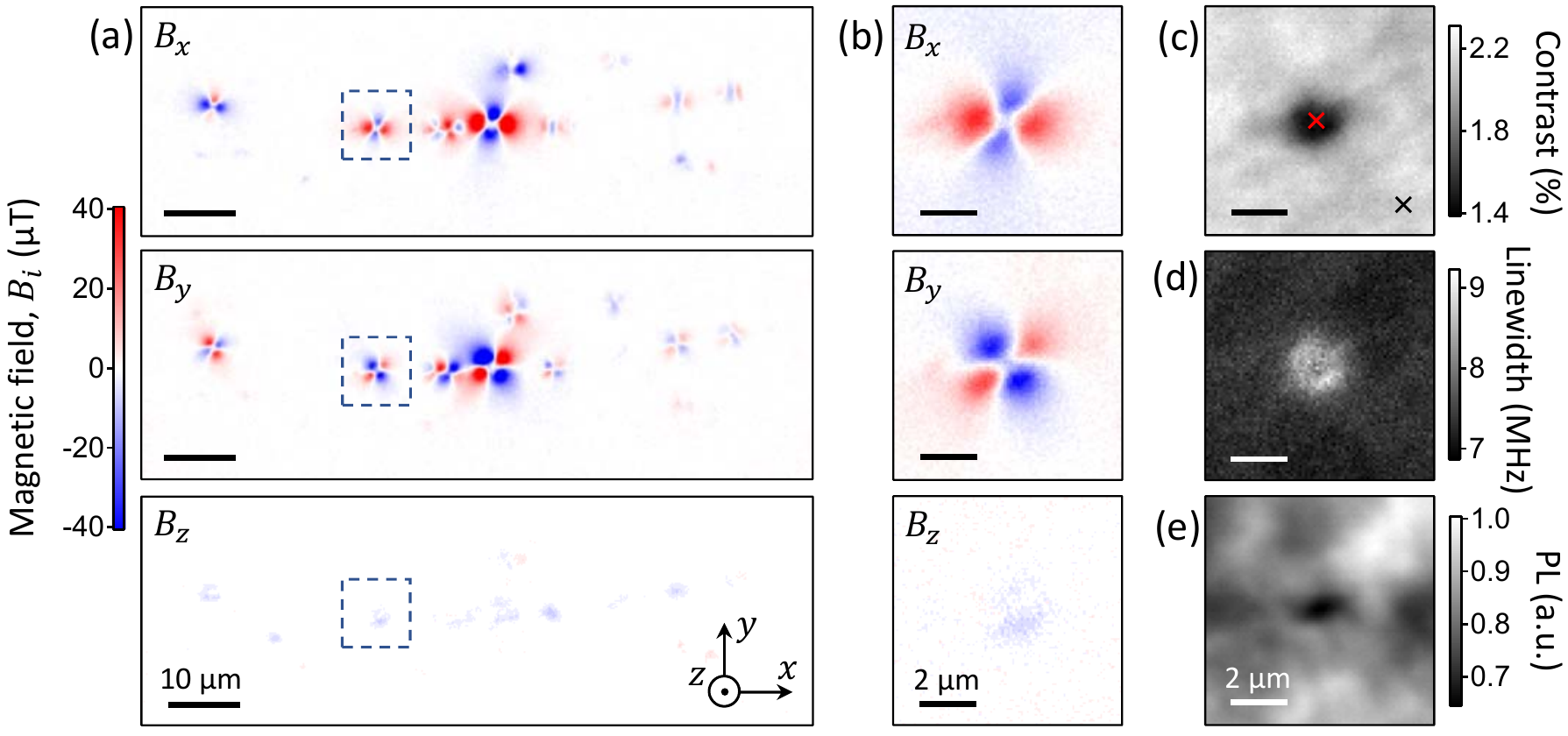}
		\caption{(a) Maps of the magnetic field components $B_x$, $B_y$ and $B_z$ (from top to bottom) showing multiple spots corresponding to magnetic nanoparticles at the diamond surface. (b) Magnetic field maps of a single ferromagnetic nanoparticle, corresponding to the dashed box drawn in (a). (c,d) Maps of the ODMR contrast (c) and linewidth (d) of the lowest-frequency resonance in the ODMR spectrum, for the same region as in (b). The red and black crosses in (c) indicate the locations of the ODMR spectra shown in Fig. \ref{Fig1}b, with matching colours. (e) Corresponding NV photoluminescence (PL) image.}
		\label{Fig2}
	\end{center}
\end{figure*} 

The results are shown in Fig. \ref{Fig2}a for a $90\times30~\mu$m$^2$ area, after subtracting the bias magnetic field ${\bf B}_0$. Several localised sources of magnetic field of various intensities can be seen, with local maxima up to $\approx100~\mu$T, corresponding to individual magnetic nanoparticles of different sizes. The magnetic field pattern looks relatively similar for all particles, with Fig. \ref{Fig2}b showing a zoom-in of a representative particle. The in-plane components ($B_x$ and $B_y$) both form a four-lobe pattern, while the out-of-plane component ($B_z$) has a much weaker intensity and its shape is not clearly defined. Such patterns are not consistent with the magnetic field produced by a single-domain nanoparticle -- roughly equivalent to a magnetic dipole -- for which all three components should be of comparable intensity regardless of the orientation of the magnetisation \cite{Coey2010}. To gain more insight into this discrepancy, Figs. \ref{Fig2}c and \ref{Fig2}d show the contrast and linewidth (defined as the full width at half maximum, FWHM) of the lowest-frequency ODMR line, respectively, for the same region as in Fig. \ref{Fig2}b. At the centre of the four-lobe magnetic field pattern, the contrast decreases from about 2.2\% to 1.5\%, while the linewidth increases from 7 MHz to 8 MHz. ODMR spectra taken on and off the centre (as indicated by crosses in Fig. \ref{Fig2}c) are shown in Fig. \ref{Fig1}b, and confirm that there is significant reduction in contrast from NV centres directly under the particle in addition to a slight line broadening. Additionally, the PL image (Fig. \ref{Fig2}e) shows a decrease in intensity ($\sim20\%$) underneath the particle.

\begin{figure*}[t!]
	\begin{center}
		\includegraphics[width=0.8\textwidth]{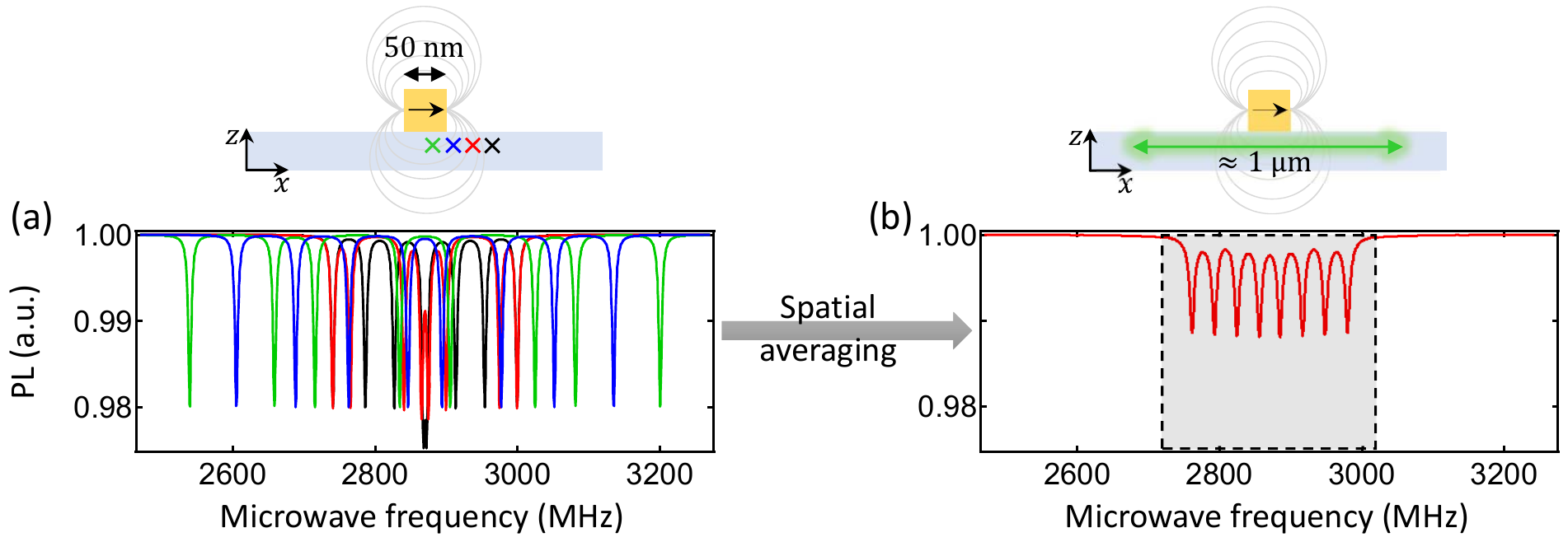}
		\caption{Simulated ODMR spectra for NV centres at a stand-off $d=20$~nm from a $50\times50\times50$~nm$^3$ particle magnetised along $x$ with $M_s=10^6$~A/m. (a) shows ODMR spectra for point-like ensembles of NV centres at different locations relative to the particle (arbitrarily chosen for illustration purpose), whereas (b) shows a spatially averaged ODMR spectrum given a $1~\mu$m optical resolution. The grey-shaded box in (b) indicates the experimentally scanned frequency window.}
		\label{Fig3}
	\end{center}
\end{figure*} 

We attribute these effects to the broad distribution of magnetic field strengths experienced by the NVs within the readout volume, which has a lateral extension in the $xy$ plane given by the optical resolution ($\approx1~\mu$m in our setup). This is illustrated in Fig. \ref{Fig3}, which shows ODMR spectra as simulated for point-like ensembles of NV centres in the vicinity of a magnetic nanoparticle (Fig. \ref{Fig3}a), as well as the spatially averaged ODMR spectrum obtained via optical readout with a near-diffraction-limited resolution of $1~\mu$m (Fig. \ref{Fig3}b). Despite the small average Zeeman shifts within the readout volume -- $100~\mu$T corresponds to a 2.8~MHz shift, much smaller than the spacing between adjacent ODMR lines -- NVs close to the source of the stray field experience larger shifts, overlapping or crossing their resonances with those from other NV orientations, or pushing them outside the probed frequency window altogether (Fig. \ref{Fig3}a). This information is irreversibly lost in the averaged ODMR (Fig. \ref{Fig3}b), leading to the observed artefacts in the reconstructed magnetic field. The purpose of the next section is to verify this interpretation through a quantitative analysis.

\section{Modelling}

\subsection{Nanoparticle with in-plane magnetisation}

\begin{figure}[t!]
	\begin{center}
		\includegraphics[width=0.4\textwidth]{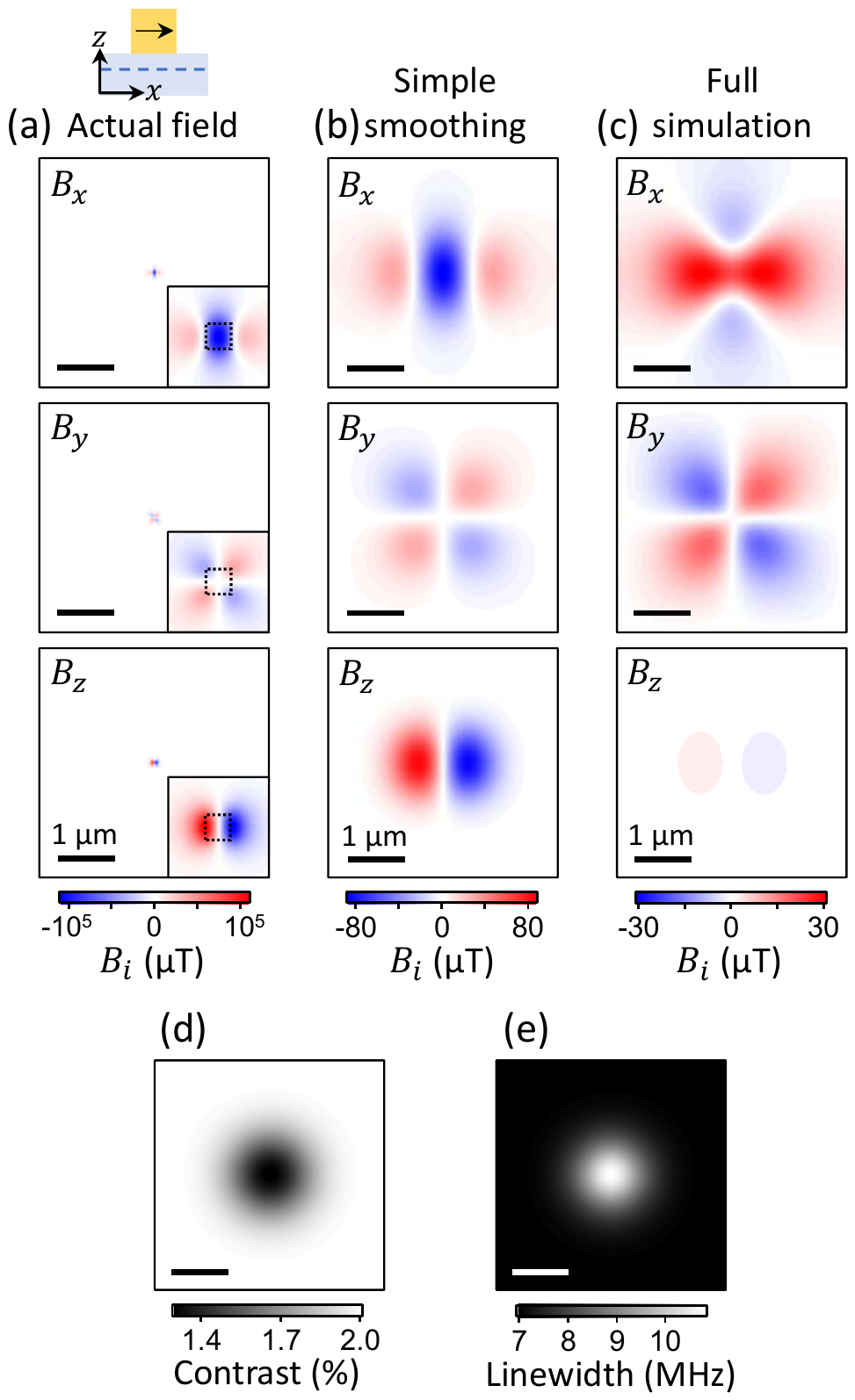}
		\caption{(a) Magnetic field calculated at a stand-off $d=20$~nm for a $50\times50\times50$~nm$^3$ particle magnetised along $x$ with $M_s=10^6$~A/m (see schematic). Inset: magnified view near the particle, whose physical footprint ($50\times50$~nm$^2$) is indicated by a dashed box. (b) Apparent magnetic field obtained by convolving the actual field (a) with a 2D Gaussian function of 1~$\mu$m FWHM. Notice the 3 orders of magnitude weaker field values in (b) compared with (a). (c) Apparent magnetic field obtained by simulating the full measurement process (see details in text). (d,e) Maps of the contrast (d) and linewidth (e) of the lowest-frequency resonance in the simulated ODMR spectrum.}
		\label{Fig4}
	\end{center}
\end{figure}

To reach a quantitative understanding of the apparent magnetic field distribution measured (Fig. \ref{Fig2}b), we modelled the measurement process including the NV response to magnetic fields, the optical readout and the spectral fitting. In the first instance, we consider a cube-shaped particle of size $50\times50\times50$~nm$^3$, magnetised along $x$ (see schematic in Fig. \ref{Fig4}a) with a saturation magnetisation $M_s=10^6$~A/m, which is typical of strong ferromagnets. The NV layer is assumed to be confined to a plane located at a distance $d=20$~nm from the diamond surface. Fig. \ref{Fig4}a shows the actual magnetic field distribution produced by the particle at the distance $d$ (see Ref. \cite{Rondin2012} for details on how this is calculated), in a $4\times4~\mu$m$^2$ area ($200\times200$~nm$^2$ in inset). The magnetic field maps feature multiple spots localised near the particle, with a FWHM of $\approx50$~nm which is consistent with the size of the particle (50 nm) and the stand-off distance $d=20$~nm. 

In our diamond, there is at least one NV centre every 30 nm (laterally) on average \cite{Simpson2016}, which should be sufficient to ensure that a few NV centres are situated near the maximum of each spot. However, the optical nature of the readout prevents distinction of NV centres separated by less than the optical resolution. While the diffraction limit is about 350 nm with our high numerical aperture objective (NA = 1.3), optical aberrations mostly due to imaging through the 30-$\mu$m-thick diamond slab deteriorates the effective optical resolution \cite{Glenn2017}, which is observed to be $\approx1~\mu$m in our setup. Therefore, the finite optical resolution is expected to smear out the magnetic field features  that are smaller than this resolution. To illustrate this effect, Fig. \ref{Fig4}b shows the simulated magnetic field maps when convolving the actual field maps (Fig. \ref{Fig4}a) with a 2D Gaussian function with a FWHM of 1~$\mu$m. The resulting magnetic field patterns are identical in shape to the actual field, but are smeared out such that the spatial extent of the field (FWHM) increases from $\approx50$~nm to $\approx1~\mu$m, and the observed maximum magnetic field is decreased from $\approx100$~mT to $\approx80~\mu$T. 

The simple Gaussian smoothing picture assumes a perfectly linear response of the NV sensors regardless of the magnetic field, neglecting their finite measurement range and other effects. One limitation is spin-state mixing due to fields larger than $\sim20$~mT, which causes the ODMR contrast to vanish unless the field is exactly aligned with the NV axis \cite{Tetienne2012}. Consequently, ODMR cannot be detected from the NVs located underneath the particle (see Fig. \ref{Fig4}a). Incidentally, the vanishing ODMR contrast is also accompanied by a reduction in PL intensity \cite{Tetienne2012}, which explains the dark spot observed in Fig. \ref{Fig2}e. Moreover, even fields under $\sim20$~mT -- which corresponds to $\sim500$~MHz Zeeman shifts -- prove challenging to measure, especially because they can assume any direction relative to the NV axes. In the magnetometry scheme employed here, optimised for fast vector magnetic field mapping, the maximum field that can be measured -- that is, the measurement range -- is $\approx0.5$~mT (i.e., $\approx15$~MHz shifts), corresponding to half the frequency spacing between two neighbouring ODMR lines (see Fig. \ref{Fig1}b). Larger fields would cause the lines to overlap or cross (see Fig. \ref{Fig3}a), preventing the vector reconstruction. We note that by performing four projective measurements with the bias field aligned along each NV axis sequentially \cite{LeSage2013,Glenn2017}, vector magnetometry could in principle be achieved over an improved range, at the expense of a significant technical overhead and a reduced sensitivity (due to the larger frequency span and multiple measurements required). With $B_0=20$~mT, for instance, magnetic fields of amplitude up to $\sim10$~mT can be measured, limited by spin mixing effects \cite{Tetienne2012}, which is still an order of magnitude lower than required to cover the full distribution calculated in Fig. \ref{Fig4}a. We also note that super-resolution optical imaging techniques can be used to overcome the diffraction limit \cite{Maurer2010}, which in principle could allow sequential readout of every single NV centre even at the current density of one NV every 30 nm. This would enable a slightly increased range ($\sim20$~mT) since no large bias field is needed, but at the cost of an enormous reduction in sensitivity because of the inherently slow acquisition rate of the technique.

As a result of this limited range, only the NV centres experiencing less than $\approx0.5~$mT contribute meaningfully to the ODMR spectrum within a given pixel using this approach. The other NVs are either outside the frequency window, or they induce additional artefacts due to line overlapping or crossing. To account for these effects, we modelled the problem as follows: (i) for each pixel of the simulated image ($5\times5$~nm$^2$), we add the bias field to the actual field generated by the particle (Fig. \ref{Fig4}a), calculate the positions of the ODMR resonances, and generate an ODMR spectrum; (ii) we apply a spatial convolution with a 2D Gaussian function (1~$\mu$m FWHM) to the ODMR maps generated in (i); (iii) for each pixel of the smoothed ODMR maps generated in (ii), we fit the ODMR spectrum with eight Lorentzian lines exactly as done to the experimental data; (iv) finally, for each pixel we convert the ODMR splittings into the magnetic field components, and subtract the bias field. The result of this process is shown in Fig. \ref{Fig4}c, and is markedly different from the actual field after smoothing (Fig. \ref{Fig4}b). In particular, the $B_x$ component exhibits a very different distribution, close to the four-lobe pattern observed experimentally. The overall shape of the other components is not significantly changed, but the field intensity shows a 1.7-fold reduction in $B_y$, and more importantly a 25-fold reduction in $B_z$. These differences can be simply understood by looking at the actual field (Fig. \ref{Fig4}a) and removing the regions where the field is above the threshold of 0.5 mT, which affects mostly the central lobe in $B_x$, as well as the two lobes in $B_z$. As can be seen by comparing Fig. \ref{Fig2}b with Fig. \ref{Fig4}c, our model reproduces well, qualitatively, the main features of the experiments, i.e. the four-lobe pattern in $B_x$, and the strongly suppressed $B_z$ component. The simulation also shows good agreement with experiment for the ODMR contrast (Fig. \ref{Fig4}d) and linewidth (Fig. \ref{Fig4}e) near the particle. We stress that we did not attempt to reach a quantitative agreement between simulation and experiment, as there are many unknown parameters, including the exact shape and size of the particle, its magnetisation, and the exact positions of each NV centre (laterally, and relative to the surface). 

\subsection{Distance dependence}

\begin{figure*}[t!]
	\begin{center}
		\includegraphics[width=1\textwidth]{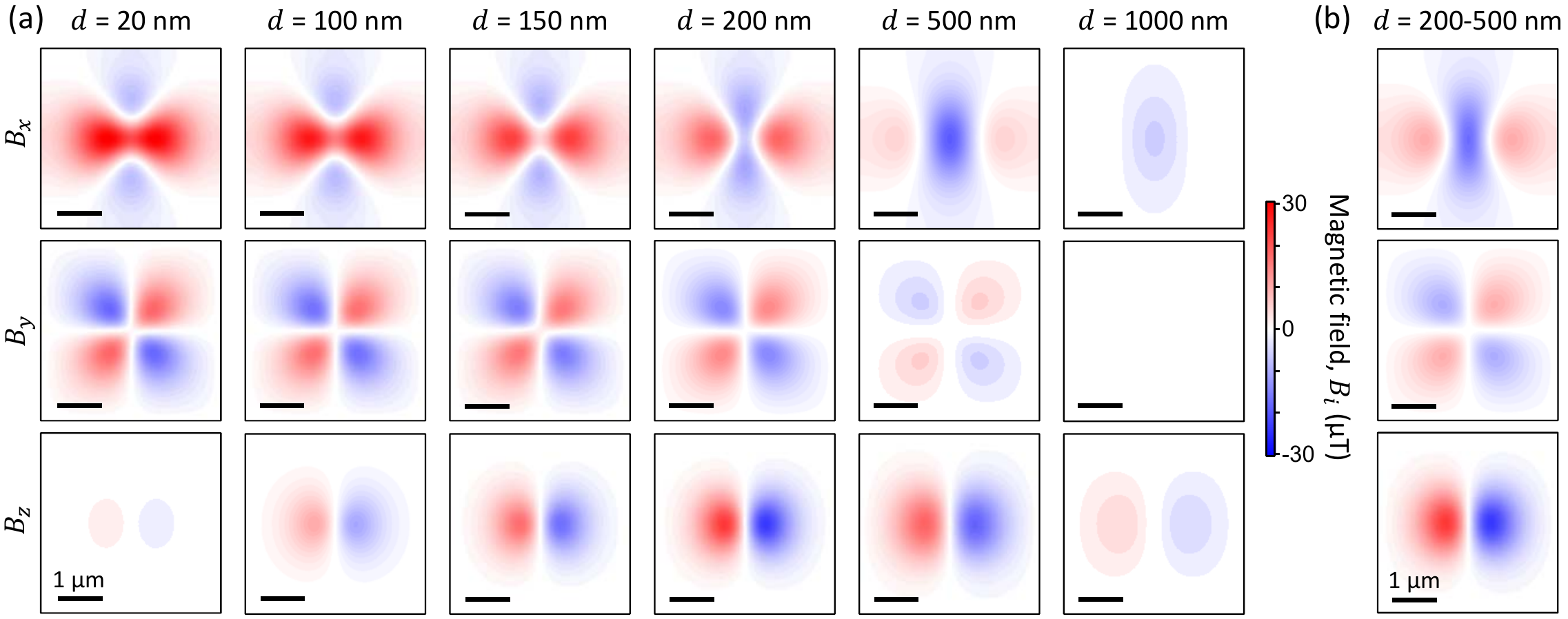}
		\caption{(a) Apparent magnetic field simulated with an increasing stand-off distance $d$ from 20 nm (far left) to 1000 nm (far right). (b) Apparent magnetic field simulated considering a uniform distribution of NV centres along $z$ between $d=200$ nm and 500 nm.}
		\label{Fig5}
	\end{center}
\end{figure*}

We have shown that the limited measurement range of the NV sensors can induce major artefacts when small magnetic objects are located very close (20 nm) to the NV layer. To avoid these artefacts, the stand-off distance between the NV layer and the target sample can be increased, for instance at $d=500$~nm the actual field from the nanoparticle is only $86~\mu$T at the maximum. This can be achieved either by adding a non-magnetic spacer layer between the diamond and the sample, or by creating the NV centres deeper into the diamond. To illustrate this, we simulated the same situation as in Fig. \ref{Fig4}, but using various distances $d$ from 20 nm up to 1~$\mu$m (Fig. \ref{Fig5}a). As the distance increases, the apparent magnetic field (as would be measured with NVs) becomes closer in shape to the actual field generated by the particle, i.e. the $B_x$ component changes from being mostly positive to mostly negative, and the $B_z$ component is recovered, with a cross-over between $d=200$ and 500 nm. For the ``artefact-free'' $B_y$, the field amplitude barely changes up to 200 nm, because this is still below the optical resolution (1~$\mu$m in these simulations). For distances above 500 nm, the field amplitude decreases because the smoothing effect becomes negligible and the field maximum then simply decays as $1/d^3$. We conclude that there is an optimum distance for strong ferromagnetic materials, typically between 200 and 500 nm, for which the artefacts are minimised and the field amplitude maximised. For weaker ferromagnetic materials or smaller particles, artefact-free imaging can be achieved at smaller distances, however, because of averaging from the near-diffraction-limited spot, decreasing $d$ below $\sim200$~nm will not provide any additional spatial information. 

In practice, the NV centres are typically created at a range of distances from the surface, i.e. $d$ is not a constant. To illustrate this, we simulated the case where the NVs are uniformly distributed between $d=200$~nm and 500 nm, which represents a good compromise as discussed above (Fig. \ref{Fig5}b). Such a distribution of depths can be readily achieved via nitrogen-doped CVD growth of diamond \citep{Kleinsasser2016}, which has the added benefit of forming NVs with improved magnetic sensitivities compared to ion implantation \cite{Tetienne2018}. We also simulated the range $d=5-20$~nm (data not shown), which is representative of the depth distributions obtained from low energy (4 keV in our experiments) nitrogen ion implantation, showing no visible differences with the $d=20$~nm case. This validates the 2D NV layer approximation used in our simulations.

\subsection{Nanoparticle with out-of-plane magnetisation}

\begin{figure*}[t!]
	\begin{center}
		\includegraphics[width=0.75\textwidth]{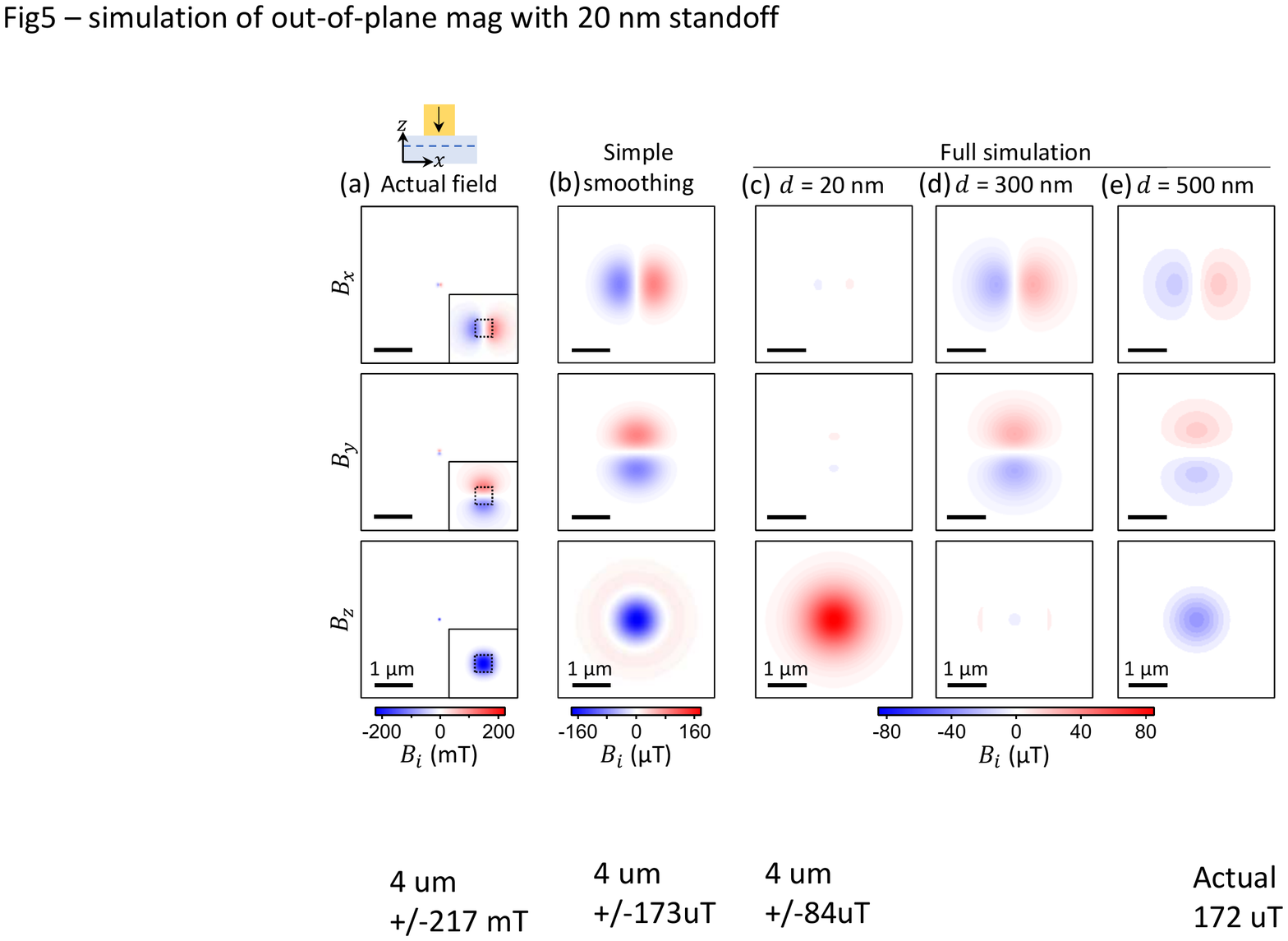}
		\caption{(a) Magnetic field calculated at a stand-off $d=20$~nm for a $50\times50\times50$~nm$^3$ particle magnetised along $z$ with $M_s=10^6$~A/m (see schematic). Inset: magnified view near the particle, whose footprint ($50\times50$~nm$^2$) is indicated by a dashed box. (b) Apparent magnetic field obtained by convolving the actual field (a) with a Gaussian spot of width 1~$\mu$m (FWHM). (c-e) Apparent magnetic field obtained by simulating the full measurement process, with $d=20$~nm (c), $d=300$~nm (d) and $d=500$~nm (e).}
		\label{Fig6}
	\end{center}
\end{figure*} 

So far, we have examined the case of a magnetic particle with a magnetisation parallel to the diamond surface (along $x$). We now consider a particle magnetised perpendicularly to the surface (along $z$), with the same parameters as in Fig. \ref{Fig4} otherwise. Figs. \ref{Fig6}a-c show the actual, smoothed and full-simulation fields as described previously, for this situation. Here, the strongest actual field is in the $B_z$ component, which reaches over 200 mT right under the particle (Fig. \ref{Fig6}a). When applying a simple Gaussian smoothing with a 1~$\mu$m FWHM (Fig. \ref{Fig6}b), the field patterns are simply smeared out, and the amplitude is decreased by three orders of magnitude, down to a maximum of $\approx170~\mu$T. Including the limited measurement range of the NV sensors, however, gives a very different result (Fig. \ref{Fig6}c). The planar components $B_x$ and $B_y$ are strongly suppressed (a 15-fold reduction relative to the simple smoothing case), and strikingly the sign of the $B_z$ component is reversed. Similar to the in-plane magnetisation case, the actual field is progressively recovered as $d$ is increased, as shown in Figs. \ref{Fig6}d and \ref{Fig6}e. Interestingly, the $B_z$ component is nearly vanishing at a cross-over distance $d=300$~nm (Fig. \ref{Fig6}d). 

Fig. \ref{Fig6} thus shows that proximity-induced artefacts can occur regardless of the direction of magnetisation in the sample, but the nature of the artefacts does depend on the direction in a characteristic manner. In fact, the strongly suppressed $B_z$ component observed in our experiments (see Fig. \ref{Fig2}a) is evidence that the magnetic particles are magnetized in the plane, while perpendicularly magnetised particles would show strongly suppressed planar components instead. In-plane magnetisation is expected for non-spherical nanoparticles deposited on a substrate, due to mechanical stability -- the particles are very likely to sit on their long axis -- and shape-induced magnetic anisotropy -- the magnetisation preferentially aligns along the long axis. 

\section{Discussion}

\begin{figure*}[t!]
	\begin{center}
		\includegraphics[width=0.9\textwidth]{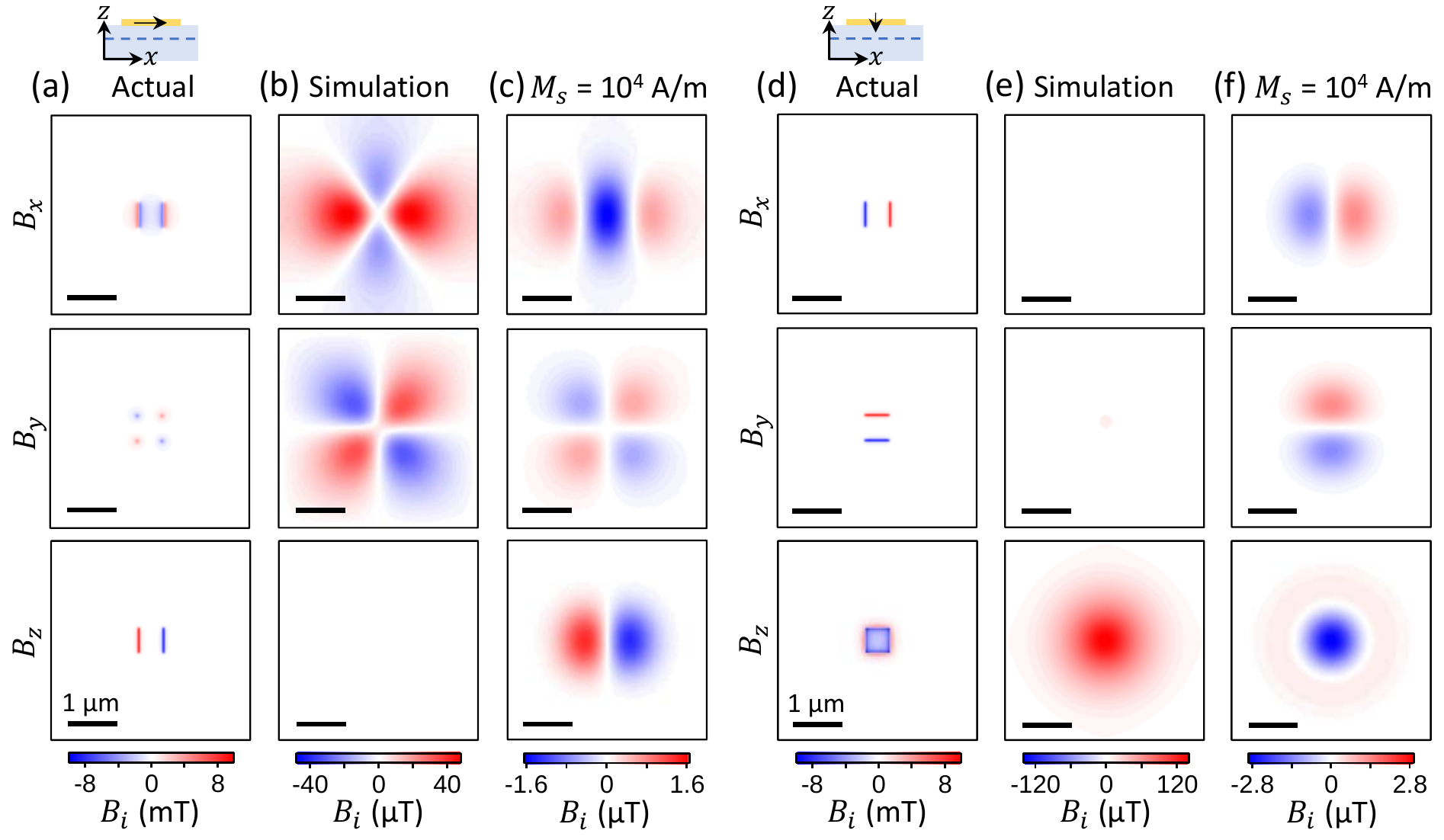}
		\caption{(a) Magnetic field calculated at a stand-off $d=20$~nm for a $500\times500\times1$~nm$^3$ flake magnetised along $x$ with $M_s=10^6$~A/m (see schematic). (b) Apparent magnetic field obtained by simulating the full measurement process, with a stand-off $d=20$~nm. (c) Same as (b) but $M_s=10^4$~A/m. (d-f) Same as (a-c) but for a flake magnetized along $z$ (see schematic in (d)).}
		\label{Fig7}
	\end{center}
\end{figure*} 

Upon examination of Figs. \ref{Fig5} and \ref{Fig6}, there seems to be no advantage from short stand-off distances such as 20 nm when probing ferromagnetic materials using a near-diffraction-limited magnetometer. Especially, because the apparent (spatially averaged) magnetic field strength does not increase below $d\approx200$~nm, the acquisition time needed to reach a specified signal-to-noise ratio remains the same. While this is true for the measurement of static magnetic fields via ODMR, other sensing modes allowed by NVs do not suffer from proximity artefacts in most situations, and on the contrary benefit greatly from a minimised stand-off distance. For instance, spectroscopic imaging of spins in samples prepared on the diamond was demonstrated via several techniques using NV ensembles \cite{Steinert2013,DeVience2015,Simpson2017}, where the signal strength falls off as $1/d^3$ for an extended object, which typically requires $d$ to be no more than $\approx20$~nm. Therefore, layers of very-near-surface NVs remain the substrate of choice whenever multi-modal imaging is desirable, and hence, accounting for ODMR-based magnetic imaging artefacts is essential.

To illustrate this point, we consider a potential application of NV sensing to ultrathin materials, i.e. with a thickness of one or a few atomic layers. In particular, an intriguing prospect is the investigation of ferromagnetism in such systems, which may be intrinsic \cite{Gong2017,Huang2017} or arise from defects or edge states \cite{Magda2014,Luxa2016,Radhakrishnan2017}. A challenge in any study of ultrathin materials is the difficulty to locate them, since they usually exhibit very low optical contrast. NV sensors offer several solutions to this problem, for instance via nuclear magnetic resonance imaging \cite{DeVience2015,Lovchinsky2017} or by using fluorescence resonance energy transfer for materials with suitable band structure, as previously demonstrated with graphene \cite{Tisler2013,Tetienne2017}. In both cases, a stand-off of order 10-20~nm at most is required to provide sufficient contrast in localising ultrathin flakes prepared on the diamond sensor, which in turn may induce artefacts when performing static magnetic imaging. 

This situation is illustrated in Fig. \ref{Fig7}, which shows magnetic field calculations for a $500\times500\times1$~nm$^3$ flake magnetized in the plane (panels a-c) or out of plane (d-f). For a strong ferromagnet ($M_s=10^6$~A/m), the field exhibits maxima near 10 mT (Figs. \ref{Fig7}a and \ref{Fig7}d). This is an order of magnitude less than in the case of the nanoparticle even though the total magnetic moment is similar in both cases, simply because the source of the field is less concentrated laterally. However, the field still largely exceeds the NV range under our measurement conditions and therefore leads to artefacts. In particular, compared to the nanoparticle case there is an even stronger suppression of the perpendicular component (planar components) for the in-plane (out-of-plane) magnetised flake (Figs. \ref{Fig7}b and \ref{Fig7}e, respectively). It is interesting to note that because the artefacts are sensitive to the local values of the actual magnetic field, they provide a pathway to distinguish different shapes and sizes of the magnetic object -- compare, e.g., Fig. \ref{Fig4}c (cubic particle) with Fig. \ref{Fig7}b (thin flake) -- while they would be essentially indistinguishable otherwise.  

As expected, the correct field patterns are recovered if the ferromagnet is sufficiently weak, for instance at $M_s=10^4$~A/m the actual field is $97~\mu$T at the maximum, well within the measurement range. The resulting apparent field reaches $\approx1.6~\mu$T for a flake magnetized in the plane (Fig. \ref{Fig7}c) and $\approx2.9~\mu$T for a flake magnetized out of the plane (Fig. \ref{Fig7}f). Such weak magnetic fields are measurable using existing NV microscopes (see Ref. \cite{Glenn2017}, see also the noise level in Fig. \ref{Fig2}b), suggesting that NV-based magnetic imaging is a viable tool to investigate ferromagnetism in ultrathin materials, such as elucidating the origin of ferromagnetism in exfoliated layered transition metal dichalcogenides \cite{Luxa2016}.

\section{Conclusion}

We showed experimentally and numerically that vector magnetic field mapping with ensembles of NV sensors can be prone to artefacts when applied to strongly ferromagnetic objects in close proximity ($<200$~nm) to the NV layer. Our modelling provides insight into the origin of these artefacts, and indicates that they can be mitigated by choosing an appropriate stand-off distance, with 500 nm being a typical optimum to also maximise the signal strength. On the other hand, the nature of these artefacts is characteristic of the direction of magnetisation, and also depends on the geometry of the magnetic object, making it a potentially useful tool for magnetic characterisation. An example application is the study of ferromagnetism in ultrathin materials, for which a small stand-off distance is desirable as it allows localisation via multi-modal imaging. This work will help researchers to choose the appropriate stand-off regime in future NV-based wide-field magnetic imaging experiments, either to minimise the proximity-induced artefacts or to exploit them.

\section*{Acknowledgements}

This work was supported in part by the Australian Research Council (ARC) under the Centre of Excellence scheme (project No. CE110001027). L.C.L.H. acknowledges the support of an ARC Laureate Fellowship (project No. FL130100119). J.-P.T. acknowledges support from the ARC through the Discovery Early Career Researcher Award scheme (DE170100129) and the University of Melbourne through an Establishment Grant and an Early Career Researcher Grant. D.A.B. and S.E.L. are supported by an Australian Government Research Training Program Scholarship. T.T. acknowledges the support of Grants-in-Aid for Scientific Research from the Ministry of Education, Culture, Sports, Science, and Technology, Japan (No. 15H03980, 26220903, and 16H06326) and Japan Science and Technology Agency (JST) CREST Grant Number JPMJCR1773, Japan.
 
\bibliographystyle{apsrev4-1}
\bibliography{bib}	
	   
\end{document}